\documentclass[12pt]{article}
\usepackage{amsfonts}

 \def\be{\begin{equation}}
 \def\ee{\end{equation}}
 \def\bea{\begin{eqnarray}}
 \def\eea{\end{eqnarray}}

\def\1{\'{\i}}                           
\def\>#1{{\bf #1}}                 
\def\d{{\rm d}}
\def\la{\lambda}
\def\curv{\mu}
\def\PP{{\cal P}}

\def\te{\theta}
\def\rr{\rho}
\def\tes{\phi}
\def\k{\kappa}

\def\SK{{\rm SK}}
\def\MSK{{\rm MSK}}

\def\jp{J_+}
\def\jm{J_-}
\def\jj{J_3}
\def\nc{{\rm nc}}
\def\cc{{\rm cc}}

\begin{document}

 \begin{center}

 {\Large{\bf{The Kepler problem on 3D spaces}}} 
\medskip

 {\Large{\bf{of  variable and constant curvature}}}
\medskip

 {\Large{\bf{from quantum algebras\footnote{Based on  the communication presented
at the Workshop in honour of Prof.~Jos\'e F. Cari\~nena {\it ``Groups, Geometry and
Physics"}, December 9--10,  2005, Zaragoza (Spain).} }}}

 \end{center}

 \medskip

 \begin{center}   {\sc  \'Angel~Ballesteros and
  Francisco~J.~Herranz}
 \end{center}

 \begin{center} {\it { 
Departamento de F\1sica,   Universidad de Burgos,\\ 09001
Burgos, Spain}}\\ e-mail:
angelb@ubu.es\quad fjherranz@ubu.es
 \end{center}

 \medskip
\medskip
\medskip

 \begin{abstract}
A quantum $sl(2,\mathbb R)$ coalgebra (with deformation parameter $z$)  is shown to
underly the construction of superintegrable Kepler potentials on 3D spaces  of variable
and constant curvature, that include the classical   spherical, hyperbolic and
(anti-)de Sitter spaces   as well as  their non-constant
curvature analogues. In this context, the non-deformed limit $z\to 0$ is identified with
the flat contraction leading to the proper Euclidean and Minkowskian spaces/potentials.
The corresponding   Hamiltonians   admit three constants of the motion coming from the
coalgebra structure. 
  Furthermore, maximal superintegrability of the Kepler potential on the spaces of
constant curvature is explicitly shown by finding an additional constant of the motion
coming from an additional symmetry that cannot be deduced from the quantum
algebra. In this way, the  Laplace--Runge--Lenz vector for such spaces is deduced and
its algebraic properties are analysed.
\end{abstract}

\newpage

%%%%%%%%%%%%%%%%%%%%%%%%%%%%%%%%%%%%%%%%%%%%%%%%%%%%%%%%%%%%%%%%%%%%%%%%%%%%%%%%%

\section{Introduction}

From the very beginning of our period as Ph.D.~students at the University of Valladolid
it has been always a pleasure for us to meet J.F.~Cari\~nena, {\em Pep\1n}. 
Since then we know that he is very interested in all the facets and approaches to the
Kepler problem (see, for instance,~\cite{car1}), and in this workshop-tribute for
his sixty-years-{\em youth}  we would like to dedicate this contribution on
the Kepler potential to {\em Pep\1n}, with our best wishes for the future.

The scheme of the paper is as follows. In the next section we show how to construct
(classical, {\em i.e.}~commutative) curved spaces by making use of quantum algebras. In
particular, we   consider the non-standard quantum deformation of $sl(2,\mathbb R)$
expressed as a deformed Poisson coalgebra and through the associated coproduct we obtain
superintegrable geodesic motions on 3D spaces of variable and constant curvature. We
are able to identify the resulting spaces with the classical spherical, Euclidean,
hyperbolic,  (anti-)de Sitter and Minkowskian  spaces and with their analogues of
non-constant curvature. In section 3 we add a Kepler potential to the former free
Hamiltonian by keeping its superintegrability. Moreover, for the spaces of constant
curvature we present a relationship between quantum algebra symmetry and Lie algebra
symmetry which, in turn, leads to additional constants of the motion. In this way, the
well-known maximal superintegrability of the Kepler potential on spaces of constant
curvature is obtained, and the explicit form of the Laplace--Runge--Lenz vector is
given for the six spaces.

%%%%%%%%%%%%%%%%%%%%%%%%%%%%%%%%%%%%%%%%%%%%%%%%%%%%%%%%%%%%%%%%%%%%%%%%%%%%%%%%%

\section{Geodesic motion on 3D curved spaces}

Let us   consider the non-standard
quantum deformation of $sl(2,\mathbb R)$ written as a Poisson coalgebra (that we denote
$sl_z(2)$) where $z$ is a {\em real} deformation parameter ($q={\rm e}^z$). The deformed 
Poisson brackets,  coproduct $\Delta$ and  Casimir ${\cal C}$ of $sl_z(2)$ are given
by~\cite{chains}:
\begin{equation}
 \{J_3,J_+\}=2 J_+ \cosh z J_-  , \quad  
\{J_3,J_-\}=-2\,\frac {\sinh zJ_-}{z} ,\quad   
\{J_-,J_+\}=4 J_3   , 
\label{ba}
\end{equation}
\begin{equation}
\begin{array}{l}
\Delta(J_-)=  J_- \otimes 1+ 1\otimes J_- ,\\[1pt]
\Delta(J_l)=J_l \otimes {\rm e}^{z J_-} + {\rm e}^{-z J_-} \otimes
J_l   ,\quad l=+,3,
\end{array}
\label{bb}
\end{equation}
\begin{equation} 
{\cal C}= \frac {\sinh zJ_-}{z} \,J_+ -J_3^2  . 
\label{bc}
\end{equation} 
A  one-particle symplectic realization of (\ref{ba}) with ${\cal C}^{(1)}=0$ reads
\begin{equation}
 J_-^{(1)}=q_1^2 ,\quad   J_+^{(1)}=
  \frac {\sinh z q_1^2}{z q_1^2}\,   p_1^2 ,\quad J_3^{(1)}=
\frac {\sinh z q_1^2}{z q_1^2 }\,    q_1 p_1  .
\label{bd}
\end{equation}
All these expressions reduce to the   $sl(2,\mathbb R)$ coalgebra under
the limit $z\to 0$, that is, the Poisson brackets and Casimir are non-deformed,
the coproduct is primitive, $\Delta(X)=X\otimes 1 + 1\otimes X$, and the
symplectic realization  is  $J_-^{(1)}=q_1^2$,  $J_+^{(1)}=  p_1^2 $ and
$J_3^{(1)}=q_1 p_1$. 

Starting from (\ref{bd}),  the coproduct (\ref{bb}) determines the corresponding
two-particle  realization and this allows one to deduce an $N$-particle realization by
applying it recursively~\cite{coalgebra}. In particular, the 3-sites coproduct,
$
\Delta^{(3)} =(\Delta \otimes \mbox{id})\circ
\Delta =(\mbox{id}\otimes \Delta )\circ \Delta $, 
gives rise to  a three-particle symplectic realization  of (\ref{ba}) defined on
$sl_z(2)\otimes sl_z(2)\otimes sl_z(2)$; namely,
\be  
\begin{array}{l}
\!\! \jm^{(3)}=q_1^2+q_2^2+q_3^2\equiv \>q^2 , \\[1pt] 
\!\!\displaystyle{ \jp^{(3)}=
  \frac {\sinh z q_1^2}{z q_1^2}  p_1^2 {\rm e}^{z q_2^2}{\rm e}^{z
q_3^2} +
 \frac {\sinh z q_2^2}{z q_2^2}  p_2^2   {\rm e}^{-z q_1^2} {\rm e}^{z
q_3^2}
 +\frac {\sinh z q_3^2}{z q_3^2}  p_3^2    {\rm e}^{-z q_1^2} {\rm
e}^{-z q_2^2}   },\\[6pt]
\!\!\displaystyle{  \jj^{(3)}=
\frac {\sinh z q_1^2}{z q_1^2 }  q_1 p_1   {\rm e}^{z q_2^2} {\rm
e}^{z q_3^2}+
\frac {\sinh z q_2^2}{z q_2^2 }  q_2 p_2   {\rm e}^{-z q_1^2}{\rm
e}^{z q_3^2} +
\frac {\sinh z q_3^2}{z q_3^2 }  q_3 p_3   {\rm e}^{-z q_1^2}{\rm
e}^{-z q_2^2}  }.
\end{array}
\label{be}
\ee
 The coalgebra approach introduced in~\cite{coalgebra} provides {\em three} functions,
coming from the two- and  three-sites coproduct of the Casimir (\ref{bc}):
\bea && 
\!\!\!\!\!\!\!\!\!\!\!\!\!\!\!\!\! 
{\cal C}^{(2)}\equiv {\cal C}_{12}   = \frac {\sinh z q_1^2 }{z q_1^2 } \,
\frac {\sinh z q_2^2}{z q_2^2} 
\left({q_1}{p_2} - {q_2}{p_1}\right)^2 {\rm e}^{-z q_1^2}{\rm e}^{z
q_2^2} ,\nonumber\\
&&
\!\!\!\!\!\!\!\!\!\!\!\!\!\!\!\!\! 
{\cal C}_{(2)}\equiv {\cal C}_{23} = \frac {\sinh z q_2^2 }{z q_2^2 } \, \frac
{\sinh z q_3^2}{z q_3^2} 
\left({q_2}{p_3} - {q_3}{p_2}\right)^2 {\rm e}^{-z q_2^2}{\rm e}^{z
q_3^2} ,\nonumber\\
&&
\!\!\!\!\!\!\!\!\!\!\!\! \!\!\!\!\!
{\cal C}^{(3)}\equiv {\cal C}_{123}  = \frac {\sinh z
q_1^2 }{z q_1^2 } \,
\frac {\sinh z q_2^2}{z q_2^2} 
\left({q_1}{p_2} - {q_2}{p_1}\right)^2 {\rm e}^{-z q_1^2}{\rm e}^{z
q_2^2} {\rm e}^{2 z q_3^2} \label{3cas}\\  &&\qquad\qquad+
\frac {\sinh z q_1^2 }{z q_1^2 } \,
\frac {\sinh z q_3^2}{z q_3^2} 
\left({q_1}{p_3} - {q_3}{p_1}\right)^2 {\rm e}^{-z q_1^2} {\rm e}^{  z
q_3^2}\nonumber  \\ &&\qquad\qquad +
\frac {\sinh z q_2^2 }{z q_2^2 } \,
\frac {\sinh z q_3^2}{z q_3^2} 
\left({q_2}{p_3} - {q_3}{p_2}\right)^2 {\rm e}^{-2z q_1^2}{\rm e}^{-z
q_2^2} {\rm e}^{z q_3^2}  .
\nonumber
\eea 
Then a large family of (minimally or weak) superintegrable Hamiltonians  can be
constructed through the following statement:
\medskip

 \noindent
{\bf Proposition 1.}
{\it 
(i) The three-particle generators (\ref{be}) fulfil the  commutation rules (\ref{ba})
with respect to the   canonical Poisson bracket 
\be
\{f,g\}=\sum_{i=1}^3\left(\frac{\partial f}{\partial q_i}
\frac{\partial g}{\partial p_i}
-\frac{\partial g}{\partial q_i} 
\frac{\partial f}{\partial p_i}\right).  
\label{bf}
\ee
(ii) These generators Poisson commute with the three   functions (\ref{3cas}).\\
(iii) Any   arbitrary function defined on (\ref{be}), ${\cal H}= {\cal
H}(\jm^{(3)},\jp^{(3)},\jj^{(3)})$ (but not on ${\cal C}$), provides a
completely integrable Hamiltonian as either  $\{{\cal C}^{(2)},{\cal
C}^{(3)},{\cal H} \}$ or $\{{\cal C}_{(2)},{\cal
C}^{(3)},{\cal H} \}$ are three functionally
independent functions in involution.\\
(iv) The   four functions $\{{\cal C}^{(2)},{\cal C}_{(2)},{\cal
C}^{(3)},{\cal H} \}$  are functionally
independent.}
 
 \medskip

As a byproduct, we obtain   superintegrable free Hamiltonians which
determine the  geodesic motion  of a particle on  certain 3D  spaces  through
\be  
\begin{array}{l} {\cal H}=\frac 12 \jp^{(3)}\, f (z\jm^{(3)} ),
\end{array}
\label{ahaa}
\ee 
where $f$ is an arbitrary  smooth function such that $\lim_{z\to
0}f(zJ_-^{(3)})=1$, so that $\lim_{z\to 0}{\cal H}=\frac
12 \>p^2$. By writing the Hamiltonian (\ref{ahaa})  as a free Lagrangian, the
  metric on the underlying 3D space can be deduced and its      sectional curvatures turn
out to be, in general, non-constant.  In this way, a quantum
deformation   can be understood as the introduction of a variable
curvature on the formerly flat Euclidean space in such a manner that the non-deformed
limit $z\to 0$ can then be identified with the {\em flat contraction} providing the proper
3D Euclidean space. Let us illustrate these ideas by recalling  two specific choices for
${\cal H}$ which have recently been studied in~\cite{plb,czec}.

%%%%%%%%%%%%%%%%%%%%%%%%%%%%%%%%%%%%%%%%%%%%%%%%%%%%%%%%%%%%%%%%%%%%%%%%%%%%%%%%%

\subsection{Spaces of non-constant curvature}

The simplest Hamiltonian (\ref{ahaa}) arises by setting $f\equiv 1$: $ {\cal H}_\nc=\frac
12
\jp^{(3)}$. This   can be rewritten
as the free Lagrangian
\be
 2{\cal T}_\nc= \frac
 {z q_1^2}{\sinh z q_1^2} \, {\rm e}^{-z q_2^2}{\rm e}^{-z q_3^2} \dot
q_1^2   +
 \frac {z q_2^2}{\sinh z q_2^2} \, {\rm e}^{z q_1^2}{\rm e}^{-z q_3^2}
\dot q_2^2    +
 \frac {z q_3^2}{\sinh z q_3^2} \, {\rm e}^{z q_1^2}{\rm e}^{z q_2^2}
\dot q_3^2  ,
 \label{ca}
\ee 
 which defines a geodesic flow on a 3D
 Riemannian space with a definite positive  metric   given  by
\be
 \d s_\nc^2=\frac {2z q_1^2}{\sinh z
 q_1^2} \, {\rm e}^{-z q_2^2}{\rm e}^{-z q_3^2} \,\d q_1^2   +
  \frac {2 z q_2^2}{\sinh z q_2^2} \, {\rm e}^{z q_1^2}{\rm e}^{-z
q_3^2}\, \d q_2^2   +
  \frac {2 z q_3^2}{\sinh z q_3^2} \, {\rm e}^{z q_1^2}{\rm e}^{z
q_2^2}\, \d q_3^2 .
 \label{ccd}
\ee The sectional curvatures $K_{ij}$ in the planes 12, 13 and 23, and the scalar
curvature $K$ turn out to be
$$  
\begin{array}{l} K_{12}=\frac z4 \,{\rm e}^{-z \>q^2}\bigl( 1+ {\rm
e}^{2 z q_3^2}- 2 {\rm e}^{2z \>q^2}\bigr) ,  \\[1pt]  K_{13}=\frac z4
\,{\rm e}^{-z \>q^2}\bigl( 2- {\rm e}^{2 z q_3^2}+ {\rm e}^{2 z
q_2^2}{\rm e}^{2 z q_3^2}- 2 {\rm e}^{2z
\>q^2}\bigr)  , \\[1pt]  K_{23}=\frac z4 \,{\rm e}^{-z \>q^2}\bigl( 2- 
{\rm e}^{2 z q_2^2}{\rm e}^{2 z q_3^2}- 2 {\rm e}^{2z
\>q^2}\bigr)  ,   \\
K=2(K_{12}+K_{13}+K_{23})=-5 z \sinh(z\>q^2) . 
\end{array}
$$ 
 
Next we introduce   new canonical coordinates
$(\rho,\te,\tes)$ and conjugated momenta $(p_\rho,p_\te,p_\tes)$ (with respect to
(\ref{bf})) defined by~\cite{czec}
\bea  &&
\cosh^2(\la_1\rho)= {\rm e}^{2z \>q^2}, \cr &&
\sinh^2(\la_1\rho)\cos^2(\la_2\te)={\rm e}^{2 z q_1^2}{\rm e}^{2 z
q_2^2}\bigl({\rm e}^{2 z q_3^2}-1 \bigr) ,\cr &&
\sinh^2(\la_1\rho)\sin^2(\la_2\te)\cos^2\tes={\rm e}^{2 z q_1^2}
\bigl({\rm e}^{2 z q_2^2}-1 \bigr),\label{xc} \\ &&
\sinh^2(\la_1\rho)\sin^2(\la_2\te)\sin^2\tes= {\rm e}^{2 z q_1^2}-1 ,
\nonumber
\eea where $z=\la_1^2$ and $\la_2\ne 0$ is an additional parameter
which can  be either a real  or a pure imaginary number~\cite{plb} and enables to deal
with Riemannian and Lorentzian signatures. Thus the metric (\ref{ccd}) is
transformed into
\be
 \d s_\nc^2=\frac {1}{\cosh(\la_1 \rr)}
 \left( \d \rr^2  +\la_2^2\,\frac{\sinh^2(\la_1 \rr)}{\la_1^2} \left( 
\d
 \te^2 + \frac{\sin^2(\la_2 \te)}{\la_2^2} \,\d\tes^2  \right) \right),
 \label{xd}
\ee which  is just the  metric of the 3D Riemannian and relativistic
spacetimes~\cite{kiev}  written in geodesic polar coordinates
multiplied by a global factor
$ {\rm e}^{-z J_-^{(3)}} \equiv {1}/{\cosh(\la_1 \rr)}$.  In the
new coordinates the sectional and scalar curvatures read
$$
  K_{12}=K_{13}= -\frac 12 \la_1^2 \,\frac{\sinh^2(\la_1
\rr)}{\cosh(\la_1
 \rr)}, \quad K_{23}=\frac{1}{2}K_{12},\quad K= -\frac 52 \la_1^2
\,\frac{\sinh^2(\la_1 \rr)}{\cosh(\la_1
 \rr)} .
$$
Therefore, according to the pair $(\la_1,\la_2)$ we have
obtained   analogues of the 3D spherical $(i,1)$,
hyperbolic   $(1,1)$, de Sitter   $(1,i)$ and anti-de Sitter $(i,i)$
spaces   with  {\em   variable} radial sectional and scalar   curvatures. These reduce
 to the flat  Euclidean $(0,1)$ and Minkowskian $(0,i)$ spaces under the limit $z\to
0$. The contraction $\la_2=0$, which is
well defined in the metric (\ref{xd}), would lead to   oscillating and
expanding Newton--Hooke $(\la_1= i,1)$   spacetimes
of non-constant curvature; again their limit $z\to 0$ would give  the   flat   Galilean
 spacetime. Nevertheless we avoid this contraction since the metric is
degenerate so that a {\em direct} relationship with a 3D Hamiltonian is lost.

The resulting
superintegrable Hamiltonian on these six curved spaces with its {\em three}
constants of motion in the latter phase space read
\be {H}_\nc=\frac 12 {\cosh(\la_1 \rr)}
 \left( p_\rr^2  + \frac{\la_1^2}{\la_2^2\sinh^2(\la_1 \rr)} \left(   
 p_\te^2 + \frac{\la_2^2}{\sin^2(\la_2 \te)}\,  p_\tes^2  \right)
\right) ,
 \label{ma}
\ee
\be {C}^{(2)}=p_\tes^2,\quad 
{C}_{(2)}=\left(\cos\tes\, p_\te-\la_2\frac{\sin\tes\, p_\tes}{\tan(\la_2\te)}  \right)^2
\!\!,
\quad {C}^{(3)}= p_\te^2+
\frac{\la_2^2\,  p_\tes^2}{\sin^2(\la_2 \te)},
\label{mb}
\ee 
where
${H}_\nc= 2{\cal H}_\nc$, ${C}^{(2)}=4{\cal C}^{(2)}$, ${C}_{(2)}= 4 \la_2^2 {\cal
C}_{(2)}$ and
${C}^{(3)}= 4 \la_2^2 {\cal C}^{(3)}$.

We remark that, in general, other choices for the Hamiltonian (\ref{ahaa}) (with $f\ne 1$)
give rise to more complicated spaces of  non-constant curvature, for which a clear
geometrical interpretation, similar to the one above developed,     remains as an
open problem. However a very particular choice of the function $\cal H$ leads to spaces
of constant curvature.

%%%%%%%%%%%%%%%%%%%%%%%%%%%%%%%%%%%%%%%%%%%%%%%%%%%%%%%%%%%%%%%%%%%%%%%%%%%%%%%%%

\subsection{Spaces of constant curvature}

If we now consider the function $f={\rm e}^{ z \jm^{(3)}}$ in (\ref{ahaa}),  we obtain
a Hamiltonian
${\cal H}_\cc=\frac 12 \jp^{(3)}
\,{\rm e}^{ z \jm^{(3)}}$ endowed  with an additional 
constant of motion ${\cal I}^{(2)}$~\cite{chains}:
\be {\cal I}^{(2)}=\frac {\sinh z q_1^2}{2 z q_1^2} \, {\rm e}^{z
q_1^2}  p_1^2  ,
\label{bjj}
\ee  
which   does not come from the
coalgebra symmetry but it is a  consequence of the
St\"ackel system~\cite{Per} defined by ${\cal H}_\cc$. Since ${\cal I}^{(2)}$ is 
functionally independent with
respect to the three previous constants of the motion (\ref{3cas}),   
  ${\cal H}_\cc$ is a {\em maximally
superintegrable}  Hamiltonian with  free Lagrangian and  associated metric 
given, in terms of (\ref{ca}) and (\ref{ccd}), by
${\cal T}_\cc = {\cal T}_\nc \,{\rm e}^{-z \>q^2}$ and
$ \d s^2_\cc = \d s^2_\nc\, {\rm e}^{-z \>q^2}$. Such a metric   is of 
Riemannian type with  constant sectional and scalar curvatures:   
$K_{ij}=z$ and  $K=6z$. 

A more familiar expression for the metric and the associated spaces can be deduced by
  applying  the change of coordinates (\ref{xc}) and next   introducing a new
radial coordinate $r$ as $\cos(\la_1 r)=1/\cosh(\la_1
\rr)$~\cite{plb}. Thus we find that $ \d s_\cc$ is  
transformed into a metric written in terms of   geodesic polar
(spherical) coordinates~\cite{kiev}:
\be
 \d s^2_\cc= \d r^2  +\la_2^2\,\frac{\sin^2(\la_1 r)}{\la_1^2}
\left(  \d
 \te^2 + \frac{\sin^2(\la_2 \te)}{\la_2^2} \,\d\tes^2  \right)   .
 \label{ye}
\ee 
According to the pair $(\la_1,\la_2)$ (we take again the simplest values: $1,0,i$), this
metric  covers     well known classical spaces of constant curvature $z=\la_1^2$: the 3D
spherical
$(1,1)$, Euclidean
$(0,1)$, hyperbolic   $(i,1)$,  anti-de Sitter   $(1,i)$, Minkowskian $(0,i)$,
 de Sitter $(i,i)$, oscillating Newton--Hooke $(1,0)$, expanding Newton--Hooke
$(i,0)$ and Galilean $(0,0)$ spaces;  we shall   avoid the non-relativistic
spacetimes with  $\la_2= 0$ as the metric is degenerate. 
Recall that  
$r$ is a radial (time-like) geodesic distance, $\te$ is either an angle in the Riemannian
spaces or a rapidity in the relativistic spacetimes ($\la_2=i/c$ with $c$ being the speed
of light), while
$\tes$ is an ordinary angle for the six spaces.

In this new phase space, the  Hamiltonian, ${H}_\cc=2{\cal H}_\cc$, reads
\be {H}_\cc=\frac 12  
 \left( p_r^2  + \frac{\la_1^2}{\la_2^2\sin^2(\la_1 r)} \left(   
 p_\te^2 + \frac{\la_2^2}{\sin^2(\la_2 \te)}\,  p_\tes^2  \right)
\right) ,
 \label{yf}
\ee 
while its {\em four} constants of motion are   ${C}^{(2)}, {C}_{(2)}, 
{C}^{(3)}$ given in (\ref{mb}) and
\be 
{I}^{(2)}=\left(\la_2\sin(\la_2\te)\sin\tes\,
p_r+\frac{\la_1\cos(\la_2\te)\sin\tes}{\tan(\la_1 r)}\,p_\te + 
\frac{\la_1\la_2\cos \tes }{\tan(\la_1 r)\sin(\la_2\te)}\,p_\tes
\right)^2 ,
\label{yg}
\ee
 where ${I}^{(2)}=4\la_2^2{\cal I}^{(2)}$.  We stress that all these results can  
alternatively be obtained by following a Lie group approach~\cite{kiev} instead of a
quantum algebra one. Explicitly, the Hamiltonian (\ref{yf})   has    a Poisson--Lie
algebra symmetry determined by a subset of $\mathbb Z_2\otimes \mathbb Z_2$ graded
contractions of $so(4)$,   $so_{\k_1,\k_2}(4)$, where $\k_i$ are two {\em  real}
contraction parameters; the six generators  $J_{\mu\nu}$
$(\mu,\nu=0,1,2,3; \mu<\nu)$ of
$so_{\k_1,\k_2}(4)$
 satisfy the following Poisson--Lie brackets:
\be
\begin{array}{lll}
\{J_{12},J_{13}\}=\k_2 J_{23},&\  \{J_{12},J_{23}\}=-J_{13},&\ 
\{J_{13},J_{23}\}=J_{12},\\[1pt]  
\{J_{12},J_{01}\}=  J_{02},&\ 
\{J_{13},J_{01}\}=J_{03},&\  \{J_{23},J_{02}\}=J_{03},\\[1pt] 
\{J_{12},J_{02}\}=-\k_2
J_{01},&\  \{J_{13},J_{03}\}=-\k_2 J_{01},&\  \{J_{23},J_{03}\}=-J_{02},\\[1pt] 
 \{J_{01},J_{02}\}= \k_1 J_{12},&\  \{J_{01},J_{03}\}= \k_1 J_{13},&\ 
\{J_{02},J_{03}\}=\k_1\k_2 J_{23},\\[1pt] 
 \{J_{01},J_{23}\}=0,&\  \{J_{02},J_{13}\}= 0,&\ 
\{J_{03},J_{12}\}=0.
\end{array}
\label{comu}
\ee
The   parameters $\k_i$ are related to the $\la_i$ through
$\k_1\equiv z=\la_1^2$ and $\k_2\equiv \la_2^2$. Consequently, the above  
six spaces of constant curvature has  a deformed coalgebra symmetry,  $\left(
sl_z(2)\otimes sl_z(2)\otimes sl_z(2)\right)_{\la_2}$, and also  a Poisson--Lie algebra
symmetry $so_{\k_1,\k_2}(4)$; the latter comprises $so(4)$ for the
spherical, $iso(3)$ for the Euclidean,
$so(3,1)$ for  the  hyperbolic, $so(2,2)$  for  
the anti-de Sitter, $iso(2,1)$ for the Minkowskian, and $so(3,1)$ for the de Sitter
space.  
 
In terms of the geodesic polar phase
space and the parameters $\la_i$ the symplectic  realization of 
the   generators $J_{\mu\nu}$   is
given by~\cite{kiev}:
\bea
&&\displaystyle{J_{01}=\cos(\la_2\theta)\,
p_r-\frac{\la_1\sin(\la_2\theta)}{\la_2\tan(\la_1 r)}\, p_\theta},\quad
 \displaystyle{J_{23}=  p_\phi},\nonumber \\
&&\displaystyle{ J_{02}=\la_2\sin(\la_2\theta)\cos\phi\, p_r+
\frac{\la_1\cos(\la_2\theta)\cos\phi}{\tan(\la_1 r)}\,
p_\theta-\frac{\la_1\la_2\sin\phi}{\tan(\la_1 r)\sin(\la_2\theta)}\,
p_\phi},\nonumber\\  
&&\displaystyle{ J_{03}=\la_2\sin(\la_2\theta)\sin\phi\, p_r+
\frac{\la_1\cos(\la_2\theta)\sin\phi}{\tan(\la_1 r)}\,
p_\theta+\frac{\la_1 \la_2\cos\phi}{\tan(\la_1 r)\sin(\la_2\theta)}\, p_\phi},\nonumber\\
&&\displaystyle{J_{12}=\cos\phi\, p_\theta-\frac{ \la_2\sin\phi}{\tan(\la_2\theta)}\,
p_\phi},\quad
\displaystyle{J_{13}=\sin\phi\, p_\theta+\frac{\la_2\cos\phi}{\tan(\la_2\theta)}\,
p_\phi} .
\label{ee}
\eea
Hence the  four constants of the motion  (\ref{mb})  and (\ref{yg})  as well as the free
Hamiltonian (\ref{yf}) are related to the generators
$J_{\mu\nu}$ through 
\bea
&&\!\!\!\!\!\!\!\!\!\!\!\!\!\!
{C}^{(2)}=J_{23}^2,\quad {C}_{(2)}=J_{12}^2,\quad 
{C}^{(3)}=J_{12}^2+J_{13}^2+\la^2_2 J_{23}^2,\quad {I}^{(2)}=J_{03}^2,\nonumber\\   
&&\!\!\!\!\!\!\!\!\!\!\!\!\!\!
2\la_2^2{H}_\cc=\la_2^2 J_{01}^2+J_{02}^2+ J_{03}^2+\la_1^2\left(
J_{12}^2+J_{13}^2+\la^2_2 J_{23}^2 \right),
\nonumber
\eea
so that ${H}_\cc$  is just the quadratic Casimir of $so_{\k_1,\k_2}(4)$
associated to the Killing--Cartan form.

%%%%%%%%%%%%%%%%%%%%%%%%%%%%%%%%%%%%%%%%%%%%%%%%%%%%%%%%%%%%%%%%%%%%%%%%%%%%%%%%%

\section{Kepler potentials}

The results of proposition 1 allows one to construct many types of  superintegrable
potentials  on 3D curved spaces through specific choices of the Hamiltonian function
${\cal H}= {\cal H}(\jm^{(3)},\jp^{(3)},\jj^{(3)})$ which could be
momenta-dependent potentials, central ones, etc. (see~\cite{jpa2D} for the 2D case). In
order to introduce a Kepler potential we consider the free Hamiltonian (\ref{ahaa}) as the
kinetic energy and  add a   term ${\cal U}(z \jm^{(3)})$ which is a smooth function such
that $\lim_{z\to 0}{\cal U}(z \jm^{(3)})=-\gamma/\sqrt{\>q^2}$ ($\gamma$ is a real
constant). Thus a family of Kepler  potentials is defined by
\be
 {\cal H}=\frac 12 \jp^{(3)}\, f (z\jm^{(3)} )+{\cal U}(z \jm^{(3)}),
\label{fa}
\ee
which can be interpeted  either as deformations of the  Kepler potential on the
flat Euclidean space (${\cal H}\to \frac 12 \>p^2-\gamma/\sqrt{\>q^2}$ when $z\to 0$), or
as   Kepler-type potentials on 3D curved  spaces. All the Hamiltonians contained within
(\ref{fa}) are {\em superintegrable}  sharing the same set of {\em three} constants of the
motion  (\ref{3cas}). 

We propose the following functions as  
the Hamiltonians containing  a Kepler potential on the 
aforementioned spaces of variable (${\cal H}^\SK_\nc$) and constant curvature
(${\cal H}^\MSK_\cc$):
\be  
\begin{array}{l} 
\displaystyle{   {\cal H}^\SK_\nc=\frac 12 \jp -
\gamma \,\sqrt{  \frac{2z}{{\rm e}^{2 z \jm}-1}} \,{\rm e}^{2 z \jm} },
 \\[4pt]
\displaystyle{ {\cal H}^\MSK_\cc=\frac 12 \jp\,{\rm e}^{ z \jm} -
\gamma \,\sqrt{  \frac{2z}{{\rm e}^{2 z \jm}-1}}   }.
\end{array}
\label{fc}
\ee 
By firstly introducing in both Hamiltonians the  symplectic realization (\ref{be}) and
secondly the new coordinates
$(\rr,\te,\tes)$ (\ref{xc}) in  $ {\cal H}^\SK_\nc$ and $(r,\te,\tes)$ in   $ {\cal
H}^\MSK_\cc$ we find that these read
\bea 
&&\!\!\!\!\!\!\! {H}^{\SK}_\nc=\frac 12 {\cosh(\la_1 \rr)}
 \left( p_\rr^2  + \frac{\la_1^2}{\la_2^2\sinh^2(\la_1 \rr)} \left(   
 p_\te^2 + \frac{\la_2^2\,  p_\tes^2}{\sin^2(\la_2 \te)}  \right)
-\frac{2\la_1 k}{\tanh(\la_1 \rr)} \right)  ,\nonumber\\
&&\!\!\!\!\!\!\!\! {H}^{\MSK}_\cc=\frac 12 
 \left( p_r^2  + \frac{\la_1^2}{\la_2^2\sin^2(\la_1 r)} \left(   
 p_\te^2 + \frac{\la_2^2\,  p_\tes^2 }{\sin^2(\la_2 \te)} \right)\right)
-\frac{\la_1 k}{\tan(\la_1 r)}  ,
\label{fe} 
\eea
where ${H}^{\SK}_\nc=2 {\cal H}^{\SK}_\nc$, ${H}^{\MSK}_\cc=2 {\cal H}^{\MSK}_\cc$ and
$k=2\sqrt{2}\,\gamma$. Hence ${H}^{\MSK}_\cc$ contains the proper Kepler
potential, either $-k/\tan r$, $-k/r$  or $-k/\tanh r$,  on six spaces of constant
curvature~\cite{car1,kiev,Schrodinger,Higgs,Leemon,6,27,
Schrodingerdual,Schrodingerdualb,18,car2}, while ${H}^{\SK}_\nc$ provides a
generalization to their  variable curvature counterpart.

The  constants of the motion   
${C}^{(2)}$ and ${C}^{(3)}$, which ensure  complete
integrability, together with the Hamiltonian allows us to 
 write  three equations, each of them depending on a canonical pair:
$$
\begin{array}{l}
\displaystyle{{C}^{(2)}(\phi,p_\phi)=p_\phi^2  ,}\quad
\displaystyle{{C}^{(3)}(\theta,p_\theta)=p_\theta^2+\frac{\la_2^2}{ 
\sin^2(\la_2\theta)}\,{C}^{(2)}
,}\\[2pt]
\displaystyle{ 
{H}^{\SK}_\nc(\rr,p_\rr)=\frac 12 {\cosh(\la_1 \rr)}
 \left(   p_\rr^2+
\frac{\la_1^2}{\la_2^2
\sinh^2(\la_1 \rr) } \,{C}^{(3)}  -\frac{2\la_1 k}{\tanh(\la_1 \rr)} \right)},\\[4pt]
\displaystyle{ {H}^{\MSK}_\cc(r,p_r)=\frac 12\, p_r^2+
\frac{\la_1^2}{2\la_2^2
\sin^2(\la_1 r) } \,{C}^{(3)} } -\frac{\la_1 k}{\tan(\la_1 r)}.
\end{array}
\label{fee}
$$
Therefore both Hamiltonians are separable and  reduced to a 1D radial system. Their
integration would lead to the solutions of such Kepler potentials; for ${H}^{\SK}_\nc$
one would find  very cumbersome   elliptic functions.

The constant of the motion ${I}^{(2)}$ (\ref{yg}) (coming from the St\"ackel system
associated to  free motion)  is lost for the Hamiltonian ${H}^{\MSK}_\cc$ defined on
the spaces of constant curvature. Nevertheless, {\em maximal superintegrability} for  
${H}^{\MSK}_\cc$ is preserved since there is an additional constant of the motion, a
component of the Laplace--Runge--Lenz vector,  which
 does not come from the coalgebra approach (as ${I}^{(2)}$)
but this is provided by the Poisson--Lie symmetry; this is a consequence of the particular
 potential expression we have considered. This property is summed up as
follows~\cite{kiev}.
 \medskip

 \noindent
{\bf Proposition 2.}
{\it Let the following three
functions   written in terms of the generators $J_{\mu\nu}$ (\ref{ee}) of
$so_{\k_1,\k_2}(4)$:
\be
\begin{array}{l}
\displaystyle{L_1=-J_{02}J_{12}-J_{03}J_{13}+k\, \la_2^2\cos(\la_2\theta),}\\[4pt]
\displaystyle{L_2=J_{01}J_{12}-J_{03}J_{23}+k\, \la_2\sin(\la_2\theta)\cos\phi,}\\[4pt]
\displaystyle{L_3=J_{01}J_{13}+J_{02}J_{23}+k\, \la_2\sin(\la_2\theta)\sin\phi.}
\end{array}
\label{hhc}
\ee
(i) The three $L_i$ Poisson commute with   ${H}^{\MSK}_\cc$ and these are the components
of the Laplace--Runge--Lenz vector on the 3D Riemannian ($\la_2$ real) and relativistic
($\la_2$ imaginary) spaces of constant curvature.\\ 
(ii) Each set  $\{{\cal C}^{(2)}=J_{23}^2,L_1,{H}^{\MSK}_\cc \}$,  $\{{\cal
C}^{(3)}-{\cal C}_{(2)}-\la_2^2{\cal C}^{(2)}=J_{13}^2,L_2,{H}^{\MSK}_\cc \}$ and 
$\{{\cal C}_{(2)}=J_{12}^2,L_3,{H}^{\MSK}_\cc \}$ is formed by three functionally
independent functions in involution.\\ (iii) The four  functions
$\{{C}^{(2)},{C}_{(2)},{C}^{(3)},{H}^{\MSK}_\cc\}$ together with any of the
components $L_i$ are functionally independent. }
 \medskip

 To end with, we present some properties satisfied by the components $L_i$.
The three generators $\{J_{12},J_{13},J_{23}\}$ span a rotation subalgebra $so(3)$ for the
three Riemannian spaces and a Lorentz one $so(2,1)$  for the three
relativistic spacetimes (see (\ref{comu})).
According to the signature of the
metric, determined by $\la_2$, the following Poisson--Lie brackets
show that the three components $L_i$ are   transformed either
as a vector under rotations when  $\la_2$ is real, or as a vector under Lorentz
transformations when  $\la_2$ is imaginary:
\be
\begin{array}{lll}
\{J_{12},L_1\}=\la^2_2 L_2,&\quad  \{J_{12},L_2\}=-L_1,&\quad 
\{J_{12},L_3\}=0,\\[1pt] 
\{J_{13},L_1\}=\la^2_2 L_3,&\quad  \{J_{13},L_2\}=0,&\quad 
\{J_{13},L_3\}=- L_1,\\[1pt] 
\{J_{23},L_1\}=0,&\quad  \{J_{23},L_2\}=  L_3,&\quad 
\{J_{23},L_3\}=- L_2 .
\end{array}
\label{fk}
\ee
The commutation rules among the components $L_i$ are found to be
\be
\{ L_i,L_j\}=2\left(\la_1^2 {C}^{(3)}- \la_2^2 {H}^{\MSK}_\cc\right) J_{ij} ,\quad
i<j,\quad i,j=1,2,3.
\label{fl}
\ee
Next  we  scale the   components as 
$\PP_1 =L_1/\la_2$, $\PP_2 =\la_2 L_2$ and $\PP_3 =\la_2 L_3$, and write
the  Poisson brackets for $J_{ij},\PP_i$ $(i,j=1,2,3)$:
\be
\begin{array}{lll}
\{J_{12},J_{13}\}=\la_2^2 J_{23},&\  \{J_{12},J_{23}\}=-J_{13},&\ 
\{J_{13},J_{23}\}=J_{12},\\[1pt]  
\{J_{12},\PP_1\}=  \PP_2,&\ 
\{J_{13},\PP_1\}=\PP_3,&\  \{J_{23},\PP_2\}=\PP_3,\\[1pt] 
\{J_{12},\PP_2\}=-\la_2^2
\PP_1,&\  \{J_{13},\PP_3\}=-\la_2^2 \PP_1,&\  \{J_{23},\PP_3\}=-\PP_2,\\[1pt] 
 \{\PP_1,\PP_2\}= \curv J_{12},&\  \{\PP_1,\PP_3\}= \curv J_{13},&\ 
\{\PP_2,\PP_3\}=\curv\la_2^2 J_{23},\\[1pt] 
 \{\PP_1,J_{23}\}=0,&\  \{\PP_2,J_{13}\}= 0,&\ 
\{\PP_3,J_{12}\}=0,
\end{array}
\label{fm}
\ee
where   $\curv=2\left(\la_1^2 {C}^{(3)}- \la_2^2 {H}^{\MSK}_\cc\right) $ is a quadratic
function on the three $J_{ij}$ through ${C}^{(3)}$ (note that ${C}^{(3)}$ does not
Poisson commute with $\PP_i$). Hence when comparing (\ref{fm}) with the Poisson--Lie
algebra  
$so_{\k_1,\k_2}(4)$ (\ref{comu}) we find that the former can be seen as a cubic
generalization of the latter under the identification  $\k_2\equiv\la_2^2$ and the
replacement of the translations $J_{0i}\to \PP_i$. The cubic Poisson brackets are those
involving two $\PP_i$ and the former contraction/deformation parameter $\k_1\equiv\la_1^2$
(the constant curvature of the space) has been replaced by the function  $\curv$.  Notice
that each set of three generators
$J_{ij},\PP_i,\PP_j$ (for $i,j$ fixed) define a cubic Higgs algebra~\cite{Higgs}.

More details on this construction as well its generalization to arbitrary dimension will 
presented elsewhere.

%%%%%%%%%%%%%%%%%%%%%%%%%%%%%%%%%%%%%%%%%%%%%%%%%%%%%%%%%%%%%%%%%%%%%%%%%%%%%%%%%

\section*{Acknowledgements}

{This work was partially supported  by the Ministerio de Educaci\'on y
Ciencia   (Spain, Project FIS2004-07913) and  by the Junta de Castilla y
Le\'on   (Spain, Project VA013C05).}

%%%%%%%%%%%%%%%%%%%%%%%%%%%%%%%%%%%%%%%%%%%%%%%%%%%%%%%%%%%%%%%%%%%%%%%%%%%%%%%%%

\end{document}